\documentclass[prl,aps,twocolumn,nopacs,superscriptaddress,nofootinbib]{revtex4-1}

\usepackage{amsmath}  \usepackage{amssymb}  \usepackage{amsfonts}  \usepackage{bm}  \usepackage{bbm}   \usepackage{braket}  \usepackage{color}  \usepackage{comment}  \usepackage{dcolumn}  \usepackage{enumerate}  \usepackage{epsfig}  \usepackage{gensymb}  \usepackage{graphicx}  \usepackage{indentfirst}  \usepackage{lmodern}  \usepackage{mathrsfs}  \usepackage{mathtools}  \usepackage{psfrag}  \usepackage{pst-all}  \usepackage{soul}  \usepackage{units}  \usepackage{xcolor}
\usepackage{upgreek}
\usepackage{float} 
\usepackage[colorlinks,linkcolor=blue,citecolor=blue,urlcolor=blue,hyperindex,driverfallback=dvipdfm]{hyperref}  \usepackage[T1]{fontenc} 

\def\ii{{\rm i}}

\begin{document} 
\def\bibsection{\section*{\refname}} 

\title{Correlative electron and light spectroscopy of strongly-coupled mid-infrared plasmon and phonon polaritons
}


\author{Pavel Gallina}
\affiliation{Central European Institute of Technology, Brno University of Technology, 612 00 Brno, Czech Republic}
\affiliation{Institute of Physical Engineering, Brno University of Technology, 616 69 Brno, Czech Republic}
\author{Andrea Kone\v{c}n\'{a}}
\email[Corresponding author: ]{andrea.konecna@vutbr.cz}
\affiliation{Central European Institute of Technology, Brno University of Technology, 612 00 Brno, Czech Republic}
\author{Ji\v{r}\'{i} Li\v{s}ka}
\affiliation{Central European Institute of Technology, Brno University of Technology, 612 00 Brno, Czech Republic}
\author{Juan Carlos Idrobo}
\affiliation{Center for Nanophase Materials Sciences, Oak Ridge National Laboratory, Oak Ridge, Tennessee 37831, USA}
\author{Tom\'{a}\v{s} \v{S}ikola}
\affiliation{Central European Institute of Technology, Brno University of Technology, 612 00 Brno, Czech Republic}
\affiliation{Institute of Physical Engineering, Brno University of Technology, 616 69 Brno, Czech Republic}




\begin{abstract}
The ability to control and modify infrared excitations in condensed matter is of both fundamental and application interests. Here we explore a system supporting low-energy excitations, in particular, mid-infrared localized plasmon modes and phonon polaritons that are tuned to be strongly coupled. We study the coupled modes by using  far-field infrared spectroscopy, state-of-the-art monochromated electron energy-loss spectroscopy, numerical simulations and analytical modeling. We demonstrate that the electron probe facilitates a precise characterization of polaritons constituting the coupled system, and enables an active control over the coupling and the resulting sample response both in frequency and space. Our work establishes a rigorous description of the spectral features observed in light- and localized electron-based spectroscopies, which can be extended to analogous optical systems with applications in heat management, and electromagnetic field concentration or nanofocusing.

\end{abstract}

\maketitle 
\date{\today} 

Polaritons are quasiparticles emerging due to strong coupling between photons and excitations in condensed matter, such as plasmons in metals and semiconductors or optical phonons in ionic crystals \cite{CLG15}. The resulting surface plasmon polaritons (PPs) and phonon polaritons (PhPs) are known to facilitate the confinement of light at the nanoscale, often deeply below the diffraction limit, which finds applications in nanoscale focusing \cite{S04,HDN08,MRB08,LLK15}, extreme waveguiding \cite{DFM14}, design of novel optical elements \cite{SDG15} or enhanced molecular detection \cite{ALD18}. Spatial confinement and energies of the polaritonic excitations can be typically tuned by nanostructuring, \textit{e.g.} in a form of gratings or the so-called optical nanoantennas \cite{H20}, but also by coupling between polaritons. Such coupling results in new hybridized modes \cite{PRH03,NOP04,CWL16,QCM16,yankovich,smith} and introduces more degrees of freedom to engineer new system functionalities and on-demand optical response \cite{KKH20,LGN21}, 

Both uncoupled and coupled polaritons in the mid-infared (MIR) energy range have been experimentally explored by far-field IR spectroscopy \cite{sikola,HNN16,PAW19,brinek}, as well as with near-field probe techniques such as scanning near-field optical microscopy (SNOM) \cite{HTK02}. Only very recently, electron energy-loss spectroscopy (EELS) \cite{E96} in a scanning transmission electron microscope (STEM) has thanks to instrumental improvements \cite{KLD14} become another suitable technique for mapping MIR polaritons with (sub-)nanometric spatial and meV spectral resolution \cite{LTH17,GKC17,KVM18,paper342,hongbin,paper359,KLE21,LBL21}. Polaritonic systems are typically analyzed by one of the aforementioned experimental techniques, however, a correlative study that would bring detailed understanding of common aspects and differences between spectral features measured by light- or electron-based spectroscopic techniques in the same sample is missing. 

In this Letter, we explore fundamental phenomena associated with nanostructured systems, where both infrared PhPs and PPs can exist. We study the electromagnetic coupling between MIR PhPs in a thin silicon dioxide film and low-energy localized surface plasmon (LSP) modes formed by the confinement of PPs in micrometer-long gold antennas. We find that far-field IR spectra can be substantially different from EEL spectra, which we confirm by experiments supported by numerical simulations and analytical modeling. We show that by precisely positioning the electron beam, the coupling between the polaritonic excitations can selectively trigger either uncoupled PhPs or coupled LSPs/PhPs. Based on our understanding of the focused-beam excitation of the polaritonic system, we also present a post-processing analysis in the EEL spectra that facilitates identification of the new hybrid modes and allows an easier comparison to the far-field optical spectra. 




\begin{figure}
\centering{\includegraphics[width=0.35\textwidth]{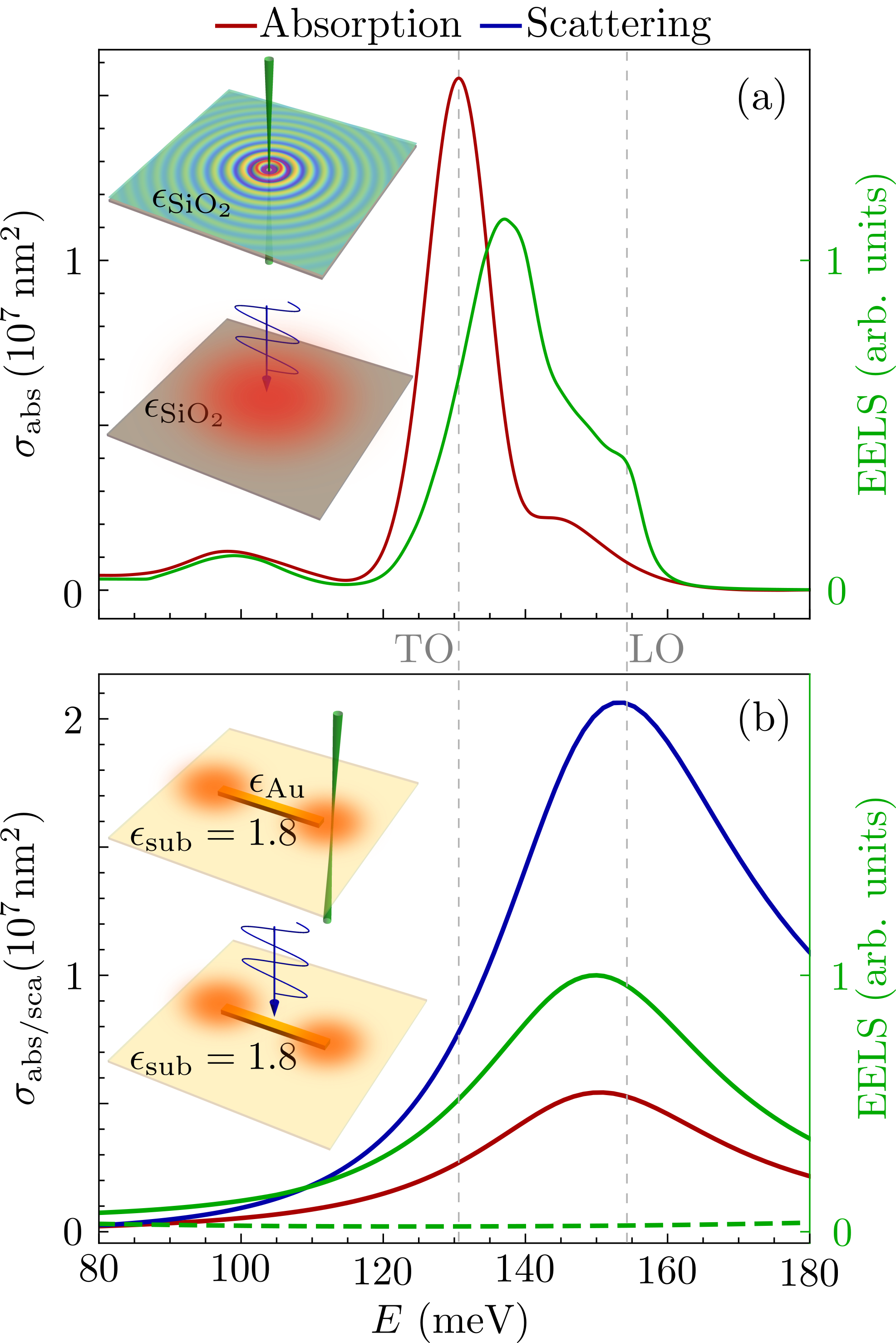}}
\caption{(a) Numerically calculated EEL (green) and optical absorption (red) spectra of an infinitely extended SiO$_2$ thin film with a thickness of $t=40$~nm. The dielectric response of SiO$_2$ is modeled using complex dielectric response obtained from experimental measurements in Ref.~\cite{KPG12}. All the spectra are probe-position invariant. Vertical dashed lines denote energies of the TO and LO phonons in SiO$_2$. (b) Spectral response of a gold (dielectric response taken from Ref.~\cite{P1985}) rectangular plasmonic antenna with the length $L=3~\upmu\mathrm{m}$, width $w=400$~nm and height $h=25$~nm on top of a substrate of the thickness $t=40$~nm with a constant dielectric response (characterized by a relative dielectric constant $\epsilon_\mathrm{sub}=1.8$). Absorption (red line) and scattering (blue line) cross sections are calculated for excitation by a plane wave impinging at normal incidence with a linear polarization aligned along the long antenna's axis (see the inset). EEL spectra are obtained for an electron beam placed 10 nm outside the antenna's corner (solid green line) and at the antenna's center (dashed green line). The electron beam energy is 60 keV in both (a,b).}
\label{Fig1}
\end{figure}

To understand the spectral response of the studied nanostructure system, we first theoretically analyze the response of its individual constituents, \textit{i.e.} a SiO$_2$ film and a long Au antenna, when they are excited by light and by a focused electron probe (as schematically shown in Fig.~\ref{Fig1}). The optical response of SiO$_2$ in the spectral region of interest is governed by a phononic mode corresponding to the Si-O-Si symmetric stretching vibration around 100~meV and another mode stemming from the Si-O-Si antisymmetric stretch around 130~meV \cite{KPG12}. The latter mode is associated with strong polarization yielding the transverse optical - longitudinal optical (TO-LO) splitting associated with the energy region, known as the Reststrahlen band (RB), where $\mathrm{Re}[ \epsilon_{\mathrm{SiO}_2}] <0$ which forbids propagation of light within bulk material.  However, in presence of boundaries, such as those imposed in the thin film geometry, new interface PhPs emerge inside the RB \cite{FK1965}. 

Due to the energy-momentum mismatch, infrared photons cannot excite PhPs in a thin film as it is demonstrated in Fig.~\ref{Fig1}(a). The most intense spectral feature corresponds to the excitation of the TO phonons which yields a strong absorption (red line). The absorption spectrum is proportional to $\mathrm{Im}[\epsilon_\mathrm{SiO_2}]$ and largely dominates over scattering (not shown). 

Focused fast electrons, on the other hand, can provide sufficient momentum and excite the interface PhPs inside the RB as demonstrated by the green spectrum in Fig.~\ref{Fig1}(a), consistent with recent experiments \cite{venkatraman2018influence,KVM18,li2019probing}. More precisely, fast electrons interacting with a thin film supporting polaritons can excite either a charge-symmetric or charge-antisymmetric coupled PhP modes \cite{LK1970,H20}. PhPs in SiO$_2$ are rather damped compared especially to those in ionic crystals \cite{HTK02,LTH17}, resulting in the two PhP modes to be spectrally indistinguishable. However, due to the symmetry of the probing field, the main peak close to 140~meV is dominated by charge-symmetric PhPs. The fast electrons can also excite the LO phonon, which requires high momenta to be activated, corresponding to the polarization along the electron trajectory. The LO phonon excitation appears as a small "shoulder" close to 155~meV.

To enable an efficient coupling between the polaritons, a spatial overlap of their electromagnetic fields as well as an overlap of their energies needs to be simultaneously targeted. As the PhPs in SiO$_2$ emerge between $\sim 130$ and 155~meV, we shall tune the LSP resonances accordingly by a careful choice of the metal used and dimensions of the plasmonic antenna. We consider gold particles of a rectangular shape and numerically simulate (see Supplementary Information) their spectral response as if the antenna were probed by a plane wave polarized along its long axis or by a perfectly focused electron beam placed close to the antenna's corner [see the insets in Fig.~~\ref{Fig1}(b)]. 

For antennas with the dimensions $3000\times 400\times 25$~nm$^3$ placed on a 40-nm thick dielectric substrate (mimicking a constant dielectric offset imposed by SiO$_2$ film), we obtain the theoretical lowest-energy dipolar LSP resonance centered around 150~meV as demonstrated in Fig.~\ref{Fig1}(b). The light excitation leads to relatively strong absorption and scattering (red and blue lines, respectively) associated with a dipolar mode that makes the electric field enhanced close to the antenna's tips. The electron beam is also capable of excitation of the same dipolar mode, which is demonstrated in the numerically calculated EEL probability (solid green line). Alternatively, focusing the electrons close to the antenna center results in a near zero signal (dashed green line) in the energy region of interest, since only higher-order modes at larger energies can be excited in this case \cite{rossouw,KNA18}. 

\begin{figure}[ht]
\includegraphics[width=0.48\textwidth]{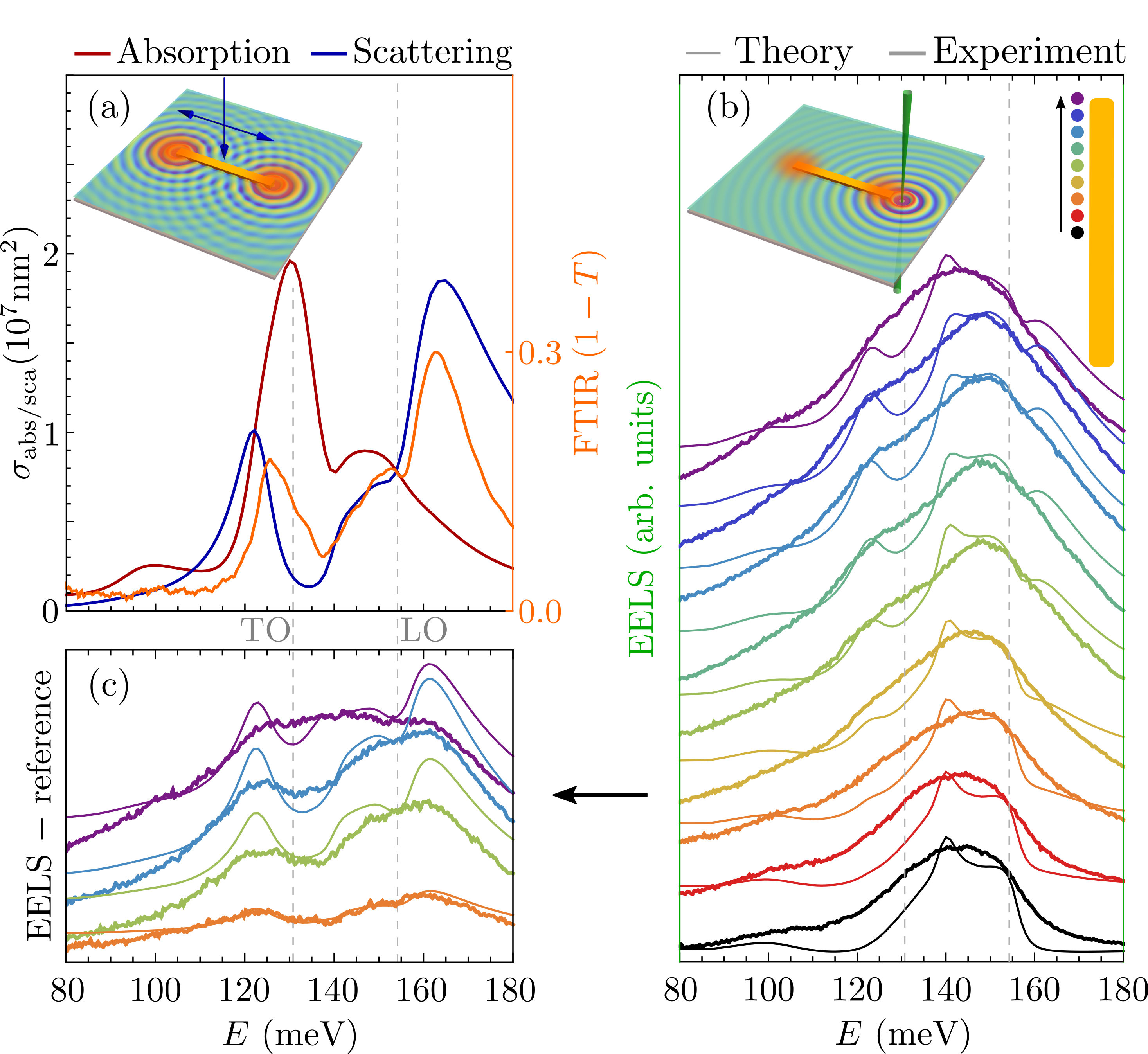}
\caption{Comparison of electron and light spectra for an approximately optimally-coupled antenna-substrate system (individual system components have the same geometry as in Fig.~\ref{Fig1}). (a) Experimentally measured FTIR spectrum corresponding to $(1-T)$, where $T$ is the total transmission (orange line) for light polarized along the long antenna axis. Numerically calculated absorption (red) and scattering (blue) cross-section spectra for an entire system are shown for comparison. (b) Experimental \textit{vs.} calculated electron spectra (thick \textit{vs.} thin lines) obtained for the 60 keV beam, which is placed just next to the antenna and scanned along the antenna's long axis. Colors approximately correspond to the electron positions as marked schematically in the inset. (c) Selected EEL spectra from (b) after subtraction of the (reference) spectrum recorded at the center of the antenna, which is dominated by uncoupled PhPs and LO phonon excitation in SiO$_2$ [black lines in (b)]. For clarity, subsequent spectra in (b,c) are vertically shifted by a constant offset. The vertical dashed lines denote energies of TO and LO phonon modes in SiO$_2$. \label{Fig2}}
\end{figure}

From the analysis of the system constituents, we can see that a setup consisting of a gold antenna on top of a thin SiO$_2$ film can sustain three dominant polaritonic modes in the energy region of interest: a single LSP mode, and symmetric and anti-symmetric thin-film PhP modes. The electromagnetic interaction between these polaritonic modes can be described by a model of three coupled oscillators captured within the matrix 
\begin{equation}
\mathbf{M}=
\begin{bmatrix}
   1/(\alpha_\mathrm{LSP}f_\mathrm{LSP}) & -K_1 & -K_2 \\
   - K_1 & 1/(\alpha_\mathrm{PhP_1}f_1) & 0 \\
    -K_2 & 0 & 1/(\alpha_\mathrm{PhP_2}f_2)
\end{bmatrix},
\end{equation}
where $\alpha_n=1/[\omega_n^2-\omega(\omega+\ii \gamma_n)]$ determines the spectral response of a mode with a resonant energy $\hbar\omega_n$, damping $\hbar\gamma_n$ and effective strength $f_n$. In the following, we assume that the PhP modes are non-radiative, while the damping of the LSP involves radiative losses, \textit{i.e.} $\gamma_\mathrm{LSP}\rightarrow\gamma_\mathrm{LSP}+\omega^2/(6\pi\epsilon_0 c^3)$ \cite{AGF10,castanie,smith}. The coupling will introduce three new hybrid modes whose eigenfrequencies and dampings are obtained from $\lvert\mathbf{M}\rvert =0$. However the coupling efficient only if both spectral and spatial overlaps of the modes' electromagnetic fields are achieved, which is described by the coupling parameters $K_1$ and $K_2$. We also assumed that PhPs do not couple to each other.

\begin{figure}[ht]
\includegraphics[width=0.48\textwidth]{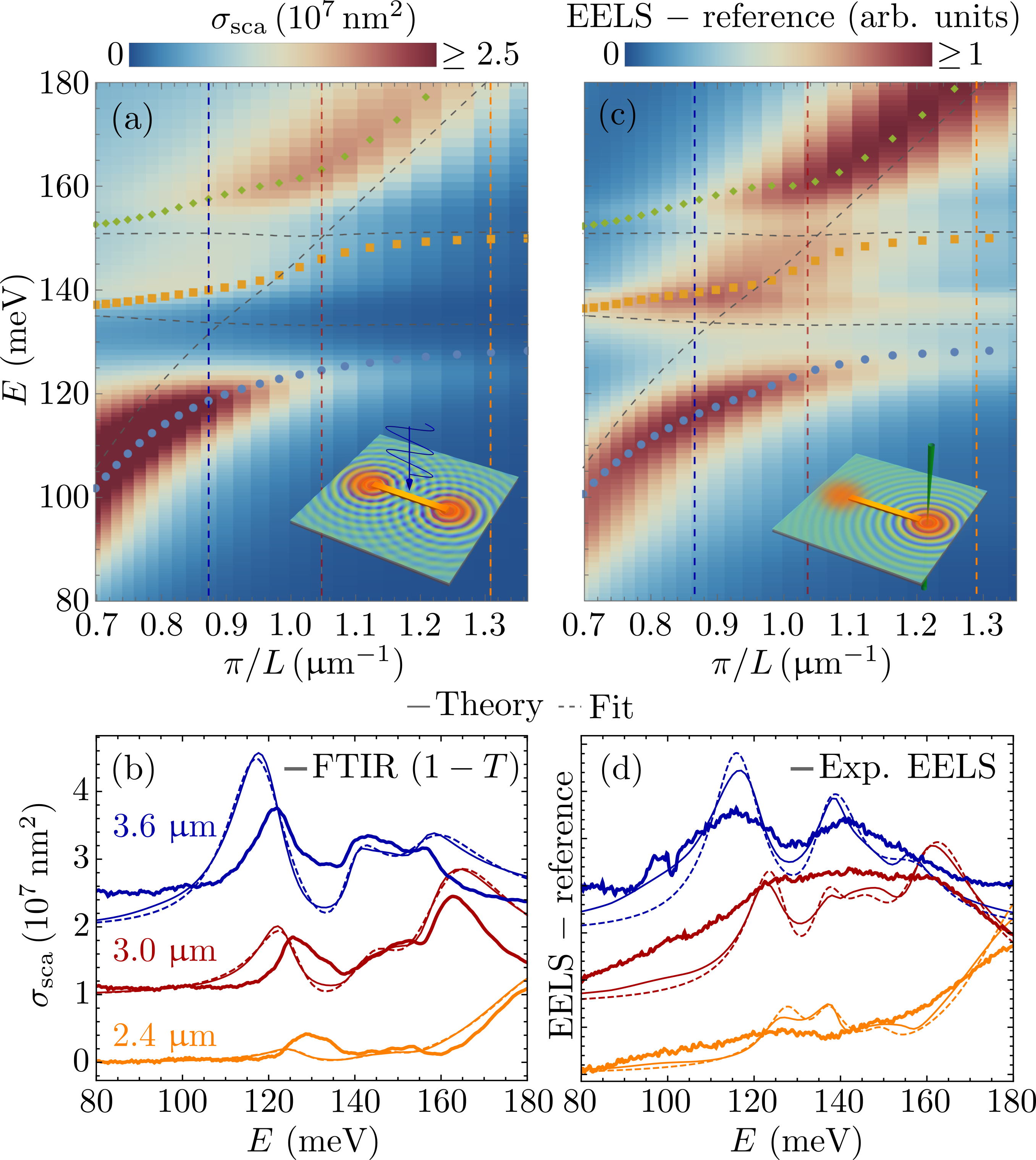}
\caption{Coupled system response as a function of the plasmonic antenna length. (a,b) Optical scattering and (c,d) reference-subtracted EEL spectra (see Fig.~\ref{Fig2} for details). Simulated pseudo-dispersions in (a,c) are obtained via the transformation $k_\mathrm{LSP}=\pi/L$, where $L$ is the antenna length and $k_\mathrm{LSP}$ is the effective wavevector of the dipolar LSP. Gray dashed lines trace energies of the uncoupled LSPs and PhPs whereas the colored symbols denote energies of the new hybrid modes characterized by eigenvalues obtained from fitting of spectra and solution of $\left\lvert \mathbf{M}\right\rvert = 0$. Examples of numerically calculated, fitted and experimental spectra (thin, dashed and thick curves, respectively) for three selected lengths (2.4/3.0/3.6 $\upmu$m) are shown in (b,d) [denoted by vertical colored dashed lines in (a,c)]. All EEL spectra were obtained for a 60 keV electron beam placed close to the corner of the antenna [violet line in Fig.~\ref{Fig2}(c)]. Experimental measurements of EELS and FTIR were performed on the same sample. \label{Fig3}}
\end{figure}

The simulated optical absorption and scattering spectra of the coupled system (red and blue lines in Fig.~\ref{Fig2}(a), respectively) exhibit several spectral features. The absorption is again dominated by the excitation of the TO phonon mode and with a similar spectral behavior as that of pure SiO$_2$ [as shown in Fig.~\ref{Fig1}(a)]. The scattering, on the other hand, clearly shows excitations beyond the RB as well as an additional weaker peak within the RB. All three peaks are associated with the new hybrid modes emerging due to the coupling. We also find that the experimental measurement of $(1-T)$, where $T$ is the transmission obtained from Fourier-transformed IR (FTIR) spectroscopy (orange line), strongly resembles the theoretically predicted scattering spectra with only a slight discrepancy in positions of the new peaks (see also Fig.~\ref{Fig3}(b) and the corresponding discussion). 

As previously mentioned, an electron beam allows to control the strength of the plasmonic excitation by simply positioning the beam at different relative positions from the antenna's center (or tip), as shown in the inset of Fig.~\ref{Fig2}(b). The corresponding simulated spectra (shown if the figure as thin lines) then capture either only non-interacting PhPs and bulk LO phonon excited in the SiO$_2$ (black), or a mixture of non-interacting PhPs and coupled LSP-PhPs (red to violet). The coupling is clearly manifested by a dip around the TO phonon position and emergence of new peaks beyond the RB. The weaker excitation inside the RB is here indistinguishable due to the presence of the uncoupled PhP and bulk signal. 

Notice that the spectrum calculated at the antenna's center is slightly different from that of a plain SiO$_2$ film shown in Fig.~\ref{Fig1}(a). This happens because the antenna represents an obstacle for the PhPs and thus favors excitation of PhPs with slightly different momenta. The experimental EEL spectra plotted in Fig.~\ref{Fig2}(b) (shown as thick lines) obtained for similar beam positions as in the simulations show less features due to limited energy resolution (between 10 and 14 meV). However, a clear broadening and emergence of shoulders of the main peak associated with the polaritonic coupling when the beam approaches the tip of the antenna can still be observed. Similar behavior of the simulated spectra is obtained when the finite experimental resolution is introduced in the simulations (see Fig.~S1 in Supplementary Information). 
In general, the far-field optical spectra of the coupled system are very different to EELS due to the absence of the spectral features corresponding to the uncoupled PhPs that are not excitable by plane waves. Interestingly, we can achieve resemblance between the light and EEL spectral signals by post-processing of EEL spectra. As only bulk LO phonon and uncoupled PhPs are excited by the beam at the center of the antenna, we take the black spectrum in Fig.~\ref{Fig3}(b) as a reference and subtract it from the spectra obtained with the beam positioned at different distances from the antenna tip. The selected reference-subtracted EEL spectra shown in Fig.~\ref{Fig2}(c) are then comparable with the optical scattering. The peak intensities and relative strength however change with the beam position, which controls the plasmon mode excitation efficiency. 

The coupling can also be adjusted by tuning the LSP energies. Fig.~\ref{Fig3} shows the dependence of the coupled system response on the parameter controlling the energy of the LSP, which is simply given by the length of the antenna's long axis $L$. Changing the antenna's dimension enables to analyze the length-dependent coupling strengths $g_{1/2}=K_{1/2}\sqrt{f_{1/2} f_\mathrm{LSP}}/\sqrt{\omega_{1/2}\omega_\mathrm{LSP}}$ \cite{novotny2010strong}, which can be obtained from fitting the optical scattering cross section and reference-subtracted EEL spectra to analytical models. We find that the overall system spectral response is approximately governed by an effective polarizability
\begin{align}
    \alpha\propto [\mathbf{M}^{-1}]_{ 11}=\frac{\alpha_\mathrm{LSP}f}{1-\sum_{n=1,2}g_n^2\omega_n\omega_\mathrm{LSP}\alpha_n\alpha_\mathrm{LSP}},
\end{align}
where $f$ is an effective response strength, and which indicates that the LSP primarily couples and decouples to the propagating IR photons as well as to the evanescent electromagnetic field supplied by the electron beam. The $[\mathbf{M}^{-1}]_{11}$ term also suggests that the PhPs are launched secondarily by the antenna. 

The scattering cross section spectra in Fig.~\ref{Fig3}(a,b) can then be modeled by \cite{AGF10,castanie}
\begin{align}
    \sigma_\mathrm{sca}\approx \frac{k^4}{6\pi\epsilon_0^2}\left\lvert\alpha\right\rvert^2,\label{Eq:scat}
\end{align}
where $\epsilon_0$ is the permittivity of vacuum and $k=\omega/c$ is the free-space wave vector of the light with photon energy $\hbar\omega$ moving at the speed of light $c$. To model the reference-subtracted EEL probability in Fig.~\ref{Fig3}(c,d), we use \cite{smith}
\begin{align}
    \mathrm{EELS}-\mathrm{reference}\approx \mathcal{F}_1(\omega)\mathrm{Im}\left\lbrace\alpha\right\rbrace+\mathcal{F}_2(\omega)\mathrm{Im}\left\lbrace \alpha_\mathrm{PhP_1}\right\rbrace,\label{Eq:EELS}
\end{align}
where $\mathcal{F}_{1/2}(\omega)=A_{1/2} \omega^{n_{1/2}}$ with $A_x$ and $n_x$ being unknown real fitting parameters. These spectral functions incorporate an overlap of the plasmonic field with the field of the electron beam. The second term in Eq.~\eqref{Eq:EELS} captures a residual spectral contribution of the uncoupled PhP, which remains even after the reference subtraction.  A residual spectral contribution remains because a slightly larger portion of PhPs (or PhPs with different momenta) can be excited when the beam is placed close to the antenna's corner. This residue can be clearly seen in Fig.~\ref{Fig3}(c) when compared to Fig.~\ref{Fig3}(a).

Fitting the simulated scattering and the reference-subtracted EEL spectra with models in Eqs.~\eqref{Eq:scat} and \eqref{Eq:EELS}, respectively, allows to obtain the parameters characterizing the uncoupled system constituents, \textit{i.e.} excitations' energies and dampings. The theoretically obtained LSP and PhP energies are interpolated by the gray dashed lines in Fig.~\ref{Fig3}(a,c). We can observe the plasmon energy linearly increasing with the inverse antenna length, which is typical for MIR plasmonic antennas on transparent substrates \cite{yueying}. Notice, however, that energies of both PhPs remain nearly constant. The equation $\left\lvert\mathbf{M}\right\rvert=0$ together with the parameters obtained from the model fitting provides the energies of the new hybrid modes, shown as colored symbols. These energies should be close to the actual peak positions, but typically do not coincide perfectly. However, we observe a close correspondence of the coupled system energies obtained for both types of probes.

The simulated spectra are compared with the experimental results for three fabricated antenna lengths in Fig.~\ref{Fig3}(b,d). The $(1-T)$ FTIR spectra exhibit a decent agreement with the calculated scattering with only slight discrepancies in the observed peak energies and relative strengths. However, it is important to keep in mind that the correspondence of the $(1-T)$ with the scattering cross section is established only empirically; an exact illumination and light collection geometry can play a role. Maybe more importantly, the optical spectra were recorded for an antenna array and thus involve many antennas with various imperfections and divergences with respect to nominal dimensions. On the other hand, each reference-subtracted experimental EEL spectrum in Fig.~\ref{Fig3}(d) was recorded for a selected antenna within the array and thus does not involve any size averaging. However, some of the fine details are hidden due to the finite instrumental resolution (see Fig.~S1 in Supplementary Information).

\begin{figure}[ht]
\includegraphics[width=0.3\textwidth]{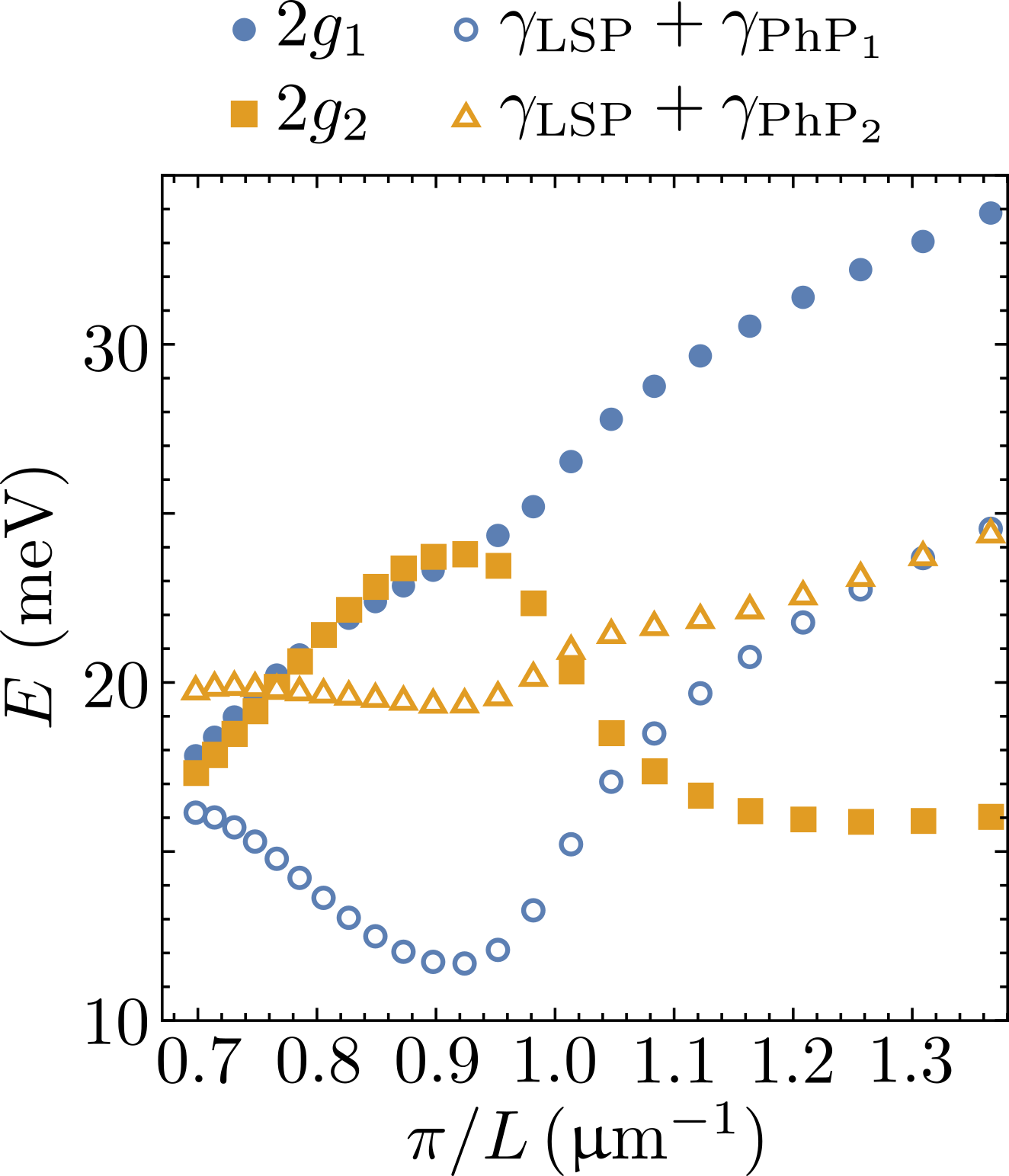}
\caption{Coupling parameters compared to their damping extracted from the fitted spectra in Fig.~\ref{Fig3}. The criterion $2g_n>\gamma_\mathrm{LSP}+\gamma_{\mathrm{PhP}_n}$ establishes the strong coupling. \label{Fig4}}
\end{figure}

The fitting enables to extract the values of the coupling strengths $g$ which are key for the classification of the coupling in the system. We theoretically predict and experimentally confirm that a rectangular Au gold nanoantena on a SiO$_2$ substrate can be in a strong coupling regime, supported by the fulfillment of the criterion \cite{torma2014strong} $2g_n>\gamma_\mathrm{LSP}+ \gamma_{\mathrm {PhP}_n}$ for coupling of the LSP with both PhPs as documented in Fig.~\ref{Fig4}, where we find optimal coupling conditions for $L\sim 3.4\,\upmu\mathrm{m}$. The fits of the optical spectra and reference-subtracted EELS provide similar coupling strengths with differences within fitting errors. Moving the electron beam towards the antenna center substantially lowers the overall efficiency of the LSP excitation, and thus the contrast of the spectral features. However, the coupling strengths stay nearly the same except for the beam positioned at the center of the antenna, where we observe a dramatic change of the coupling strengths towards zero as the dipolar LSP cannot be excited anymore. 


In conclusion, here we have performed a comparative experimental study of spectra of the coupled antenna-substrate system obtained with far-field light and near-field electron spectroscopy. The study reveals fundamental differences when probing a complex polaritonic system with light and focused electron probes. We show that a precise positioning of the electron beam offers the possibility to probe coupled or uncoupled excitations at will, thus offering complementary information to that obtained from far-field optical spectroscopy. We also present a post-processing analysis in the EEL spectra, that consists of subtracting a reference from spectra acquired for different beam positions with respect to the nanostructure system to reveal the strength of coupling between phonon and plasmon polariton excitations. The workflow method presented here can be generalized for the study of excitations arising when geometry, topology and different materials are used to generate new hybrid optical systems.

\section*{ACKNOWLEDGEMENTS} 
We acknowledge the support by the Czech Science Foundation (Grant No.*20-28573S*), European Commission (H2020-Twininning project No. 810626 – SINNCE, M-ERA NET HYSUCAP/TACR-TH71020004),*BUT* – specific research No.*FSI-S-20-648*5, and Ministry of Education, Youth and Sports of the Czech Republic (CzechNanoLab Research Infrastructure – LM2018110). J.C.I. acknowledges support by the Center for Nanophase Materials Sciences (CNMS), which is a U.S. Department of Energy, Office of Science User Facility.  Preliminary experimental work at ORNL was conducted, in part, using instrumentation within ORNL’s Materials Characterization Core provided by UT-Batelle, LLC, under Contract No. DE-AC05-00OR22725 with the U.S. Department of Energy, and sponsored by the Laboratory Directed Research and Development Program of Oak Ridge National Laboratory, managed by UT-Battelle, LLC, for the U.S. Department of Energy. We thank our colleagues Michal Hor\'{a}k, Marek Va\v{n}atka and Vojt\v{e}ch \v{S}varc for their advice on lithography on TEM membranes.

\section*{COPYRIGHT NOTICE}
This manuscript has been authored by UT-Battelle, LLC under Contract No. DE-
AC05-00OR22725  with  the  U.S.  Department  of  Energy. The  United  States  Government 
retains and the publisher, by accepting the article for publication, acknowledges that the 
United  States  Government  retains  a  non-exclusive,  paid-up,  irrevocable,  world-wide 
license to publish or reproduce the published form of this manuscript, or allow others to 
do  so,  for  United  States  Government  purposes.  The  Department  of  Energy  will  provide 
public access to these results of federally sponsored research in accordance with the DOE 
Public Access Plan (http://energy.gov/downloads/doe-public-access-plan). 





\newpage
\begin{widetext}
\section*{Supplementary Information} 
\setcounter{figure}{0}
\renewcommand{\thefigure}{S\arabic{figure}}
\subsection{Numerical simulations}
Finite-difference time-domain simulations of far-field optical spectra were obtained using the Lumerical software~\cite{lumerical}. A single rectangular antenna placed on a semi-infinite membrane was illuminated by a linearly polarized plane wave provided by a total-field scattered-field (TFSF) source. The scattering (absorption) spectra were calculated from the scattered (total) power flux monitors placed outside (inside) the TFSF source. The whole simulation domain with dimensions of 10~$\upmu$m $\times$ 10~$\upmu$m $\times$ 6~$\upmu\mathrm{m}^3$ was enclosed in a perfectly matched layer.

EEL spectra were obtained using the finite element method implemented within the Comsol Multiphysics software \cite{comsol}, where we calculate the induced electromagnetic field emerging in the interation of the nanostructure with a line current representing the focused electron probe. The EEL probability is then evaluated as \cite{paper149}
\begin{align}
    \Gamma(\omega)=\frac{e}{\pi\hbar\omega}\int\mathrm{d}z\,E^\mathrm{ind}_z(\mathbf{R}_\mathrm{b},z,\omega)\,\mathrm{e}^{-\mathrm{i}\omega z/v},\label{Eq:Gamma}
\end{align}
where $e$ is the elementary charge, $\hbar\omega$ energy, and $v$ is the electron velocity. $z$ denotes the optical axis along which the fast electron propagates, $\mathbf{R}_\mathrm{b}$ is the impact parameter (i.e. position in the transverse plane with respect to the optical axis which the electron trajectory intersects) and where we integrate the $z$ component of the induced electric field along the beam trajectory. 

\subsection{Experimental details}
\subsubsection{Fabrication}
Electron beam lithography on SiO$_2$ TEM membranes (thickness 40~nm) was performed using scanning electron microscopy Mira3 (Tescan) with the laser interferometry stage (Raith). Subsequent gold deposition was done using electron beam evaporator (Bestec).

\subsubsection{FTIR}
Fourier-transform infrared spectroscopy was measured using an IR microscope [Vertex 70v and an IR microscope Hyperion 3000 (Bruker)] with an aperture allowing the signal collection from an area of 50~$\upmu$m $\times$ 50~$\upmu\mathrm{m}^2$ in a spectral range of 600-6000 cm$^{-1}$ and resolution of 2 cm$^{-1}$.

\subsubsection{STEM-EELS}
Electron energy-loss spectra were acquired using a Nion monochromated aberration-corrected scanning transmission electron microscope (MAC-STEM) instrument operated at 60 kV accelerating voltage \cite{KLD14,hachtel}. The measurements were performed with a convergence semiangle of 30 mrad, a collection semiangle of 20 mrad, a beam current of $\sim 20$~pA, using a Nion Iris spectrometer with a dispersion of 0.4 meV/channel \cite{lovejoy}, and an energy resolution (the full width at half-maximum of the zero loss peak) was between 10 meV to 14 meV. 

\subsection{Influence of finite resolution in EELS}

Figure \ref{FigS1} shows a comparison of the theoretical EEL spectra when is convoluted with a Gaussian function of 14~meV FWHM to mimic the energy resolution of the involved STEM-EELS setup.
\begin{figure*}[ht]
\includegraphics[width=0.7\textwidth]{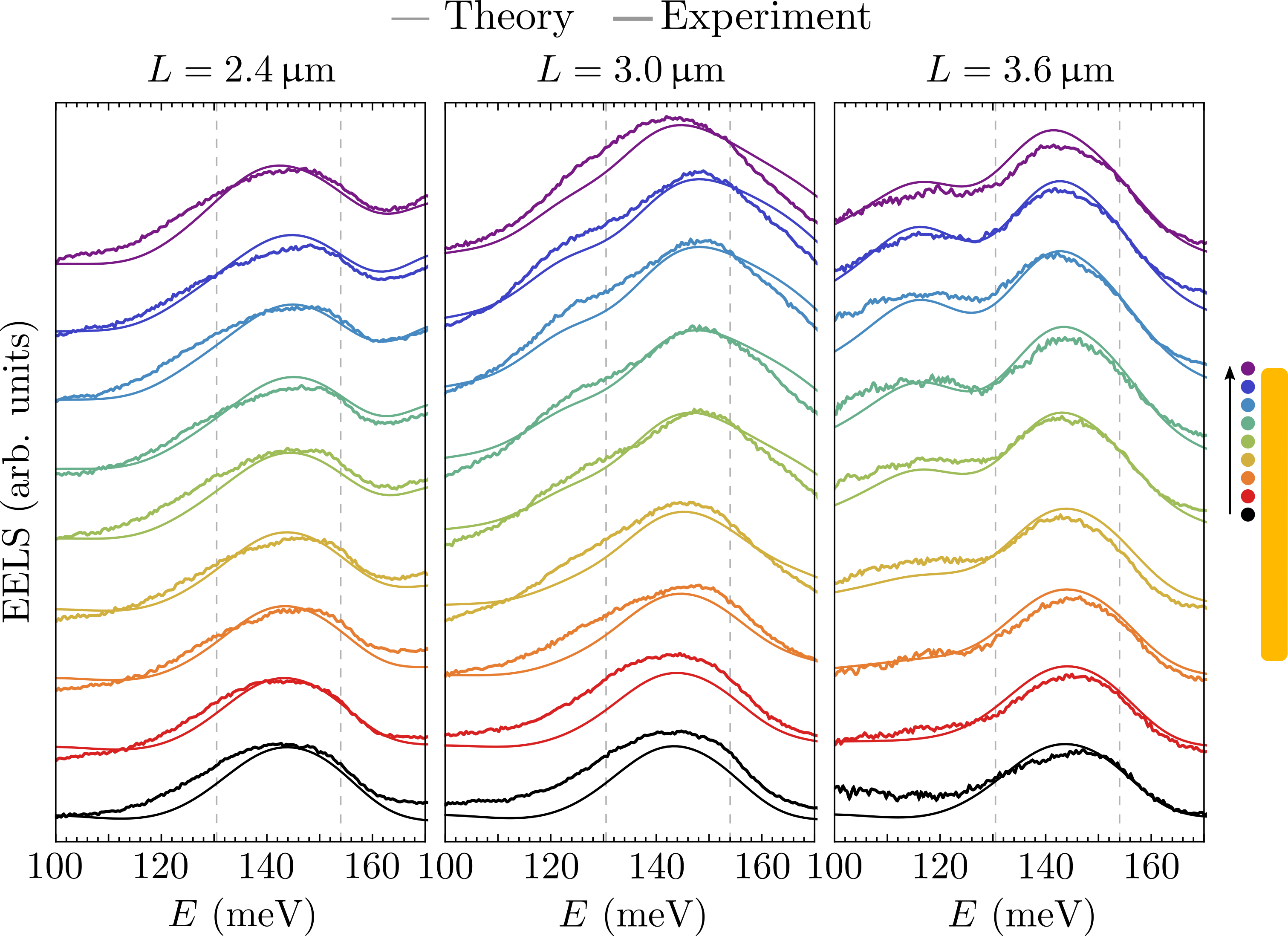}
\caption{Comparison of theoretically calculated EEL spectra after convolution with a Gaussian function of 14~meV FWHM (thin lines) with experimentally measured spectra (thick lines) for the varying beam position as shown in the inset and three antenna's lengths. \label{FigS1}}
\end{figure*}
\end{widetext}





\bibliography{refsL.bib,Coupling_EELS.bib} 

\end{document}